%% file: dblock.tex
\documentstyle[cite,epsf]{europhys}
\input{euromacr}

\makeatletter
\def\@maketitle{\rm\vbox to0pt{}\vskip-34pt
 \parindent\z@
 \def\EPLogo{Preprint\hfill March 31, 1999}
 \EPLogo
 \vglue 50pt
 {\Large\bf \@title\par\smallskip}
 \vskip 16pt \leftskip 25pt
 {\normalsize\@author }
 \vskip  4pt {\normalsize\it\@institute}
 \vskip 16pt {\small\@pacs}
 \vskip 36pt
 \setcounter{page}{\value{startpage}}}
\makeatother

\makeatletter
\def\lesssim{\mathrel{\mathpalette\vereq<}}
\def\vereq#1#2{\lower3pt\vbox{\baselineskip1.5pt \lineskip1.5pt
\ialign{$\m@th#1\hfill##\hfil$\crcr#2\crcr\sim\crcr}}}
\makeatother

\newcommand{\mc}[1]{\multicolumn{1}{c}{#1}}
\newcommand{\figurewidth}{100mm}

\begin{document}

\euro{0}{0}{0}{0}
\Date{}

\title{Do crossover functions depend on the shape\\
       of the interaction profile?}
\author{Erik Luijten\inst{1,2} \And Kurt Binder\inst{2}}
\institute{%
        \inst{1} Max-Planck-Institut f\"ur Polymerforschung, Postfach 3148,
                 D-55021 Mainz, Germany \\
        \inst{2} Institut f\"ur Physik, WA 331,
                 Johannes Gutenberg-Universit\"at, D-55099 Mainz, Germany}
\rec{}{}
\pacs{%
    \Pacs{05}{70.Jk}{Critical point phenomena}
    \Pacs{64}{60.Fr}{Equilibrium properties near critical points, critical
       exponents}
}

\shorttitle{Do crossover functions depend on the shape of the interaction
            profile?}

\maketitle

\begin{abstract}
  We examine the crossover from classical to non-classical critical behaviour
  in two-dimensional systems with a one-component order parameter. Since the
  degree of universality of the corresponding crossover functions is still
  subject to debate, we try to induce non-universal effects by adding
  interactions with a second length scale.  Although the crossover functions
  clearly depend on the range of the interactions, they turn out to be
  remarkably robust against further variation of the interaction profile.  In
  particular, we find that the earlier observed non-monotonic crossover of the
  effective susceptibility exponent occurs for several qualitatively different
  shapes of this profile.
\end{abstract}

\section{Introduction}
In recent years, there has been a revived interest in the nature of the
crossover from classical to non-classical (asymptotic) critical behaviour upon
approach of the critical point.  This crossover between two universality
classes can be observed in a great variety of many-body systems, including pure
fluids, polymer mixtures and micellar solutions, and is driven by the ratio
between the reduced temperature $t \equiv (T-T_{\rm c})/T_{\rm c}$ (where
$T_{\rm c}$ is the critical temperature) and a system-dependent parameter~$G$,
the Ginzburg number. The dependence of observables on this ratio is described
by so-called {\em crossover functions}. Compared to our knowledge of critical
exponents, for which very accurate, consistent estimates are available from
renormalization-group~(RG) calculations, series expansions, experiments and
numerical calculations, the situation for non-asymptotic critical phenomena is
not so clear-cut.  Most theoretical predictions for crossover functions are
obtained by means of RG-based methods. Examples include the work of Nicoll and
Bhattacharjee~\cite{nicoll81}, who used an RG-matching method to calculate
crossover functions for the one-phase region ($T>T_{\rm c}$) to second order in
$\varepsilon=4-d$ ($d$ denotes the spatial dimensionality), and of Bagnuls and
Bervillier~\cite{bagnuls85}, who applied massive field theory in $d=3$. The
latter work was then extended to the two-phase region as well~\cite{bagnuls87},
although only for temperatures relatively close to~$T_{\rm c}$. Very recently,
the approach of Ref.~\cite{bagnuls87} was used to calculate the full crossover
function for the susceptibility exponent below~$T_{\rm c}$~\cite{pelissetto98}.
A more phenomenological approach has been taken by Belyakov and
Kiselev~\cite{belyakov92}, who presented a generalization of first-order
$\varepsilon$-expansions. Although the different methods vary in mathematical
rigour, they all suggest that the crossover functions are {\em universal\/}
functions of the ratio~$t/G$, under the additional restriction $t \to 0$, $G
\to 0$. This limit, which is referred to as {\em critical\/} crossover, implies
that one must consider the limit in which the coefficient~$u$ of the quartic
term in the Landau--Ginzburg--Wilson~(LGW) Hamiltonian goes to zero (as can be
realized, {\em e.g.}, in systems with a diverging interaction
range~\cite{medran}).  Experimental systems evidently do not obey these
restrictions: Here $G$ is a fixed parameter and the crossover functions are
obtained by varying $t$, where it is generally assumed that within the critical
region, {\em i.e.\/}\ for $t$ sufficiently small, one still observes a
universal crossover.  Recent work~\cite{anisimov95,chi3d,anisimov98} has shown
that this assumption is only partially correct.  In this paper, we therefore
examine the role of some of the parameters that might be held responsible for
deviations from the field-theoretic crossover curves and show what degree of
universality one may still expect.

Anisimov {\em et al.}~\cite{anisimov92} have suggested that, apart from the
correlation length~$\xi$, an additional (mesoscopic) length scale may determine
the nature of the crossover behaviour for complex fluids. Based upon earlier
work~\cite{chen90b}, a corresponding parametric crossover function was proposed
and in Refs.~\cite{anisimov95,melnichenko97} it was shown that this function of
{\em two\/} variables can indeed describe the crossover of the susceptibility
exponent for several experimental systems displaying qualitatively different
behaviour.  On the other hand, there are quite a number of experimental results
that can be described in terms of the above-mentioned {\em single}-parameter
functions, although it should be noted that only few experiments have yielded
accurate results for the effective exponents, which are defined as the
logarithmic derivatives of the crossover functions~\cite{kouvel64}.
Furthermore, most experiments only partially cover the crossover region, which
occupies several decades in the reduced temperature.  Thanks to recent
algorithmic developments, numerical methods can circumvent both of these
limitations in an efficient way, which has already led to several notable
results~\cite{chicross,cross,chi3d}. In particular, it was demonstrated in
Ref.~\cite{chi3d} that the crossover function for the effective susceptibility
exponent $\gamma_{\rm eff}^{+}$ (pertaining to $T>T_{\rm c}$) for
three-dimensional systems with a {\em finite\/} interaction range~$R$ is
steeper than the functions presented in Refs.~\cite{bagnuls85,belyakov92}.  The
reason for this discrepancy lies in $u$ not being small for small~$R$.  Whereas
the {\em scale\/} of the crossover is determined by the Ginzburg number, the
{\em shape\/} of the crossover functions is determined by
$u$~\cite{anisimov98}. This makes it difficult to obtain an explicit expression
for the crossover.  One possibility is to invoke the description of
Ref.~\cite{anisimov95}.  A (still somewhat phenomenological) fit to this
description is indeed possible~\cite{anisimov98}, but a further demonstration
that the crossover functions depend on more than one parameter is clearly
desirable.  The numerical results presented in
Refs.~\cite{chicross,cross,chi3d} have been obtained for a block-shaped
interaction profile (the so-called equivalent-neighbour model), where the
interaction strength is kept constant within a radius~$R_m$ and zero beyond
that. As long as the interaction has a finite range, different interaction
profiles will lead to the same universal properties~\cite{medran}, but not
necessarily to the same crossover functions.  Thus, there is a twofold
objective in studying the effect of a modified interaction profile. In the
first place, we want to study the effect of introducing an additional length
scale in the block-shaped interaction profile, in order to study its influence
on the crossover functions.  Secondly, this modification of the interaction
profile allows us to examine, on a more general level, the dependence of
crossover functions on the shape of the interaction profile and thus to shed
some light on the universal nature of these functions.

\begin{figure}
\leavevmode \centering \epsfxsize 8cm
\epsfbox{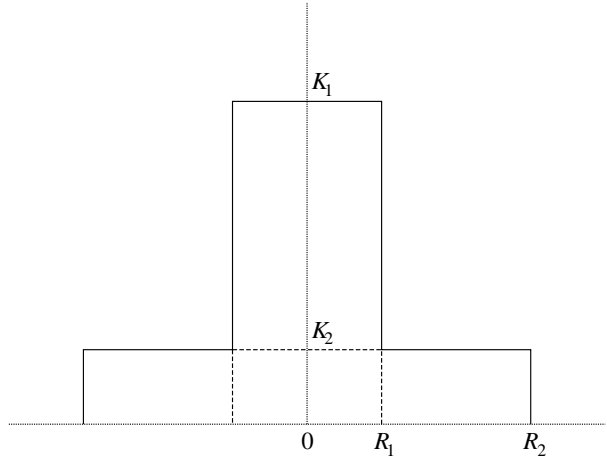}
\caption{The interaction profile studied in this work. Each spin interacts
with a ferromagnetic coupling~$K_1$ with all its neighbours within a
distance~$R_1$ and with a coupling~$K_2$ with those at a distance between $R_1$
and~$R_2$.}
\label{fig:profile}
\end{figure}

\begin{table}[b]
\caption{Some properties of the interaction profiles studied in this work. The
systems are listed in order of increasing~$R$, except for the last one, which
is the only profile for which $R_1 < R_{\rm min}$ ($R_1^2=27$ is the borderline
case corresponding to the minimum in~$R$).}
\label{tab:ranges}
\centering
\renewcommand{\arraystretch}{1.2}
\begin{tabular}{rrrrrrlll}
\hline
\mc{$R_1^2$} & \mc{$R_2^2$} & \mc{$R^2$} & \mc{$z_1$} & \mc{$z_2$} & 
\mc{$\alpha z_1/z_2$} & \mc{$K_{\rm c}$} & \mc{$Q$} &
\mc{$y_{\rm h}$} \\ 
\hline
  2 &  10 &  2.54 &   8 &  28 &  4.57 & 0.00872817~(2)    & 0.8559~(3) &
  1.8744~(9)  \\
  6 &  32 &  6.60 &  20 &  80 &  4.00 & 0.002918960~(7)   & 0.8566~(8) & 
  1.8739~(14) \\
 25 &  32 & 13.38 &  80 &  20 & 64.00 & 0.000836683~(2)   & 0.8551~(8) &
  1.8738~(13) \\
 27 & 140 & 28.06 &  88 & 348 &  4.05 & 0.000598920~(2)   & 0.856~(2)  &
  1.873~(2)   \\
 49 & 140 & 31.41 & 148 & 288 &  8.22 & 0.0003934150~(11) & 0.858~(3)  &
  1.872~(6)   \\
 70 & 140 & 39.32 & 220 & 216 & 16.30 & 0.0002775163~(9)  & 0.856~(5)  &
  1.873~(3)   \\
 93 & 140 & 48.80 & 292 & 144 & 32.44 & 0.0002140278~(7)  & 0.856~(5)  &
  1.878~(4)   \\
114 & 140 & 57.91 & 356 &  80 & 71.20 & 0.0001777575~(4)  & 0.850~(4)  &
  1.867~(5)   \\
  4 & 140 & 49.99 &  12 & 424 &  0.45 & 0.001683755~(5)   & 0.859~(4)  &
  1.873~(5)   \\
\hline
\end{tabular}
\end{table}

\section{Simulational aspects and determination of critical properties}
In order to maximize the numerical sensitivity to variations in the crossover
function, we have restricted ourselves to the two-dimensional~(2D) case only.
An additional attractive aspect of this system is, that the effective
susceptibility exponent~$\gamma_{\rm eff}^{-}$ exhibits a remarkable
non-monotonicity~\cite{chicross}. It is well possible that such a peculiar
property is particularly sensitive to variations in the interaction profile.
Thus, we have carried out Monte Carlo simulations for a 2D Ising model, defined
on a square lattice of size $L \times L$ with periodic boundary conditions.
Each spin interacts with a strength~$K_1$ with all its neighbours within a
distance~$r \leq R_1$ (domain~$D_1$) and with a strength~$K_2$ with all its
neighbours within a distance~$R_1 < r \leq R_2$ (domain~$D_2$). This means that
the block-shaped interaction profile of Ref.~\cite{medran} has been generalized
to the double-blocked case depicted in Fig.~\ref{fig:profile}. The strength
ratio $K_1/K_2$ is denoted by the parameter~$\alpha$, which throughout this
work is supposed to be greater than unity. In order to suppress lattice
effects, all range dependences are expressed in terms of the effective
interaction range~$R$, defined by
\begin{equation}
 R^2 \equiv \frac{\sum_{i \neq j} |{\bf r}_i-{\bf r}_j|^2 K_{ij}}%
            {\sum_{i \neq j} K_{ij}} \;,
\end{equation}
which for our interaction profile reduces to $R^2 = (\alpha \sum_{i \in D_1}
r_i^2 + \sum_{i \in D_2} r_i^2)/z_{\rm eff}$, where $z_{\rm eff} \equiv \alpha
z_1 + z_2$, with $z_i$ the number of neighbours in domain~$D_i$.  The strength
ratio was chosen as $\alpha=16$, in order to create a strong asymmetry between
the two domains. The value of $R_2$ was kept fixed at~$\sqrt{140}$. In the
finite-size scaling analyses, the minimum system size has to be of the order
of~$R_2^2$ and a maximum linear system size $L=1000$ thus implies that the
results cover a factor~$7$ in~$L$. The effective interaction range~$R$ was then
varied by varying~$R_1$. Both for $R_1 \to 0$ and for $R_1 \to R_2$, $R$ will
take its maximum value (which in the continuum limit approaches $R_2/\sqrt{2}$)
and it will reach a minimum at $R_1 = R_{\rm min}$. In the continuum limit, the
corresponding effective range is $R^2 = R_{\rm min}^2 =
R_2^2/(\sqrt{\alpha}+1)$. Although the same values for~$R$ can be reached with
$R_1 < R_{\rm min}$ and $R_1 > R_{\rm min}$, it should be noted that the two
cases greatly differ. For example, for $R_2^2=140$ and $\alpha=16$ (where
$R_{\rm min}^2=28.06$ is reached for $z_1=88$, {\em i.e.\/}\ $26 \leq R_1^2
\leq 28$) one may obtain $R^2 \approx 50$ by choosing either $R_1^2=4$ or
$R_1^2=93$, but in the former case $D_1$ contains 12 out of~436 interacting
neighbours, compared to 292 out of~436 in the latter case.  This means that the
integrated coupling ratio $\alpha z_1/z_2$, which indicates the relative
contribution of the two domains to the total integrated coupling, is 0.45 in
the first case and 32 in the second case. So, in combination with the original
block-shaped profile, we can realize three qualitatively different interaction
profiles and study the dependence of the crossover curve on the profile.
Table~\ref{tab:ranges} lists some properties of the interaction profiles
considered in this work. The three profiles with $R_2^2 < 140$ were added in
order to reach very small effective interaction ranges as well.  For each
choice, we have carried out extensive simulations using a dedicated cluster
algorithm for long-range interactions~\cite{lr-alg}.  The critical properties
of each individual system were determined via finite-size scaling analyses,
along the lines described in Ref.~\cite{medran}. The critical coupling, for
which an accurate value is required to attain the proper crossover curve, has
been determined from the amplitude ratio $Q = \langle m^2 \rangle^2 / \langle
m^4 \rangle$ and we have obtained the magnetic exponent $y_{\rm h}$ from the
absolute magnetization density $\langle |m| \rangle$ (see
Table~\ref{tab:ranges}).  One notes that for all systems $Q$ is in good
agreement with the 2D Ising value $Q_{\rm I} \approx 0.856216$~\cite{q-2d} and
$y_{\rm h}$ lies very close to~$15/8$. This confirms the expectation that all
systems belong to the 2D Ising universality class. In addition, we have
determined the magnetic susceptibility for $t<0$ from the fluctuation relation
$\chi = L^d (\langle m^2 \rangle - \langle |m|^2 \rangle) / k_{\rm B}T$ and the
critical finite-size amplitudes of $\langle |m| \rangle$ and $\langle m^2
\rangle$.

\begin{figure}
\leavevmode \centering \epsfxsize\figurewidth
\epsfbox{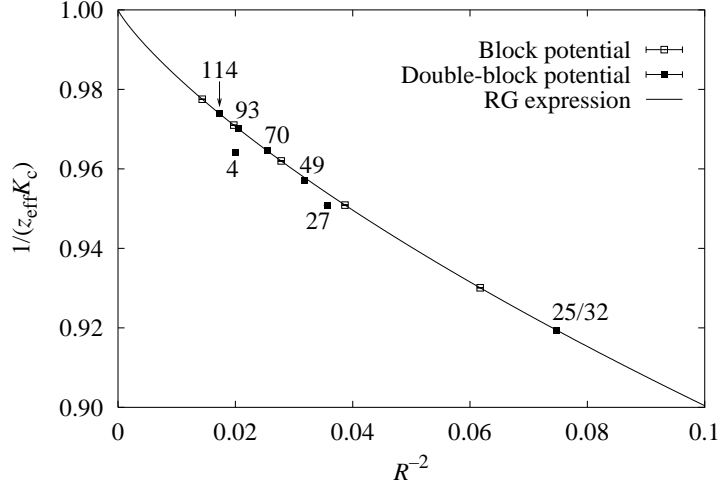}
\caption{The critical temperatures of the various models as a function of the
interaction range. The open squares and the curve refer to the block-shaped
profile of Ref.~\protect\cite{medran} and the black squares to the profiles
studied in the present paper. The numbers indicate the value of the
parameter~$R_1^2$ (with $R_2^2=140$ fixed); the indication $25/32$ refers to
the system with $R_1^2=25$ and~$R_2^2=32$.}
\label{fig:kc}
\end{figure}

\section{Range dependence of critical properties and analysis of the crossover
functions} Figure~\ref{fig:kc} shows the critical temperatures as a function of
the interaction range. All temperatures are expressed in units of the critical
temperature of the mean-field model, $T_{\rm c}=1/( z_{\rm eff} K_{\rm c} )$.
Since fluctuations are less suppressed when the interaction range decreases,
one observes that $T_{\rm c}$ is gradually depressed for smaller~$R$. More
importantly, the figure illustrates that a definitely non-universal quantity
like $T_{\rm c}$ does not depend on the effective interaction range alone:
Although for several systems the critical temperature lies on the curve
describing $T_{\rm c} (R)$ for the block-shaped interaction profile, the
systems with $R_1 \lesssim R_{\rm min}$ ($R_1^2=4,27$) exhibit a clear
deviation from this curve. Apparently, the latter systems show (for a
given~$R$) the greatest deviation from the mean-field model, in the sense that
the suppression of fluctuations is least efficient.

\begin{figure}[t]
\leavevmode \centering \epsfxsize\figurewidth
\epsfbox{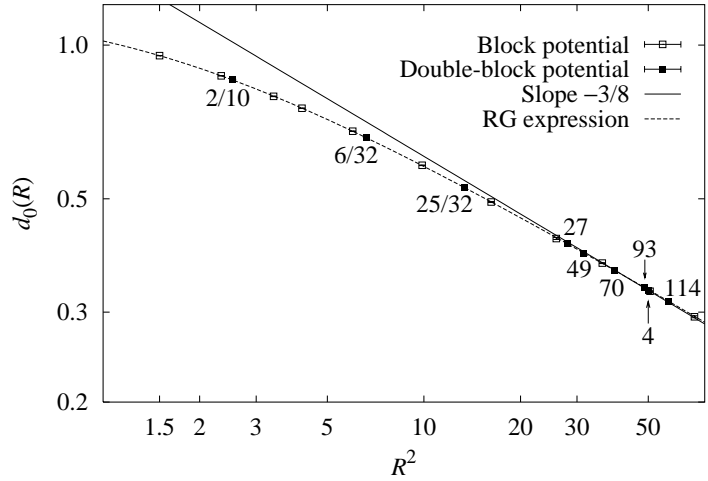}
\caption{Critical finite-size amplitude of the absolute magnetization
density. The dashed curve indicates the RG~expression fitted to the open
squares; clearly, it also describes the black squares (referring to the
interaction profiles of the present work) very well. The numbers are explained
in the caption of Fig.~\protect\ref{fig:kc}.}
\label{fig:mabs}
\end{figure}

In contrast, no such deviations are observed for the critical finite-size
amplitude of, {\em e.g.}, the absolute magnetization density. This quantity,
defined as $d_0 \equiv \lim_{L \to \infty} L^{d-y_{\rm h}} \langle |m|
\rangle$, has an asymptotic range dependence proportional to $R^{(3d-4y_{\rm
h})/(4-d)}$~\cite{medran}. It turns out that the systems investigated in this
work do not only follow this asymptotic law, but also for smaller~$R$ agree
very well with the range dependence found for the block-shaped profile, see
Fig.~\ref{fig:mabs}. This even holds for highly asymmetric profiles such as
$R_1^2=4;R_2^2=140$.

\begin{figure}
\leavevmode \centering \epsfxsize\figurewidth
\epsfbox{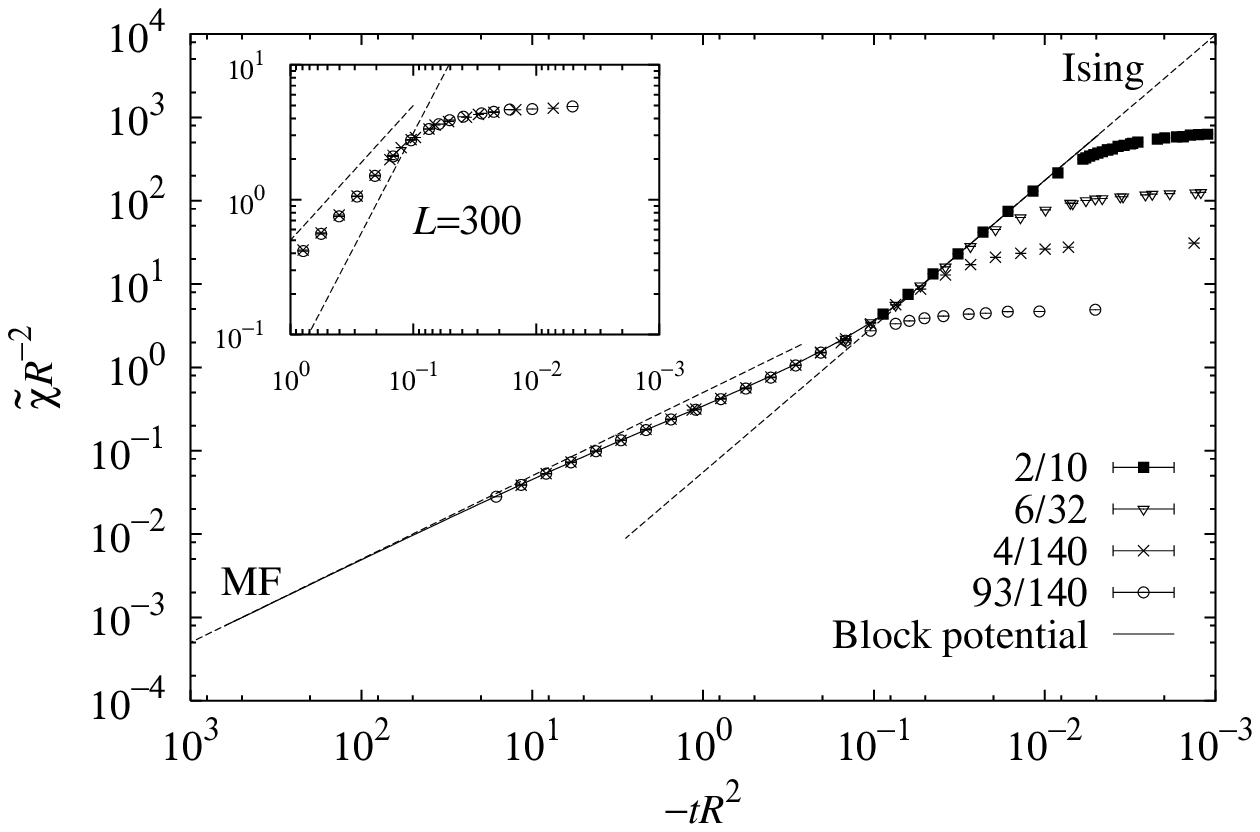}
\caption{The crossover function for the connected susceptibility below~$T_{\rm
c}$. $\tilde{\chi}=\chi/C(R)$, where $C(R)$ is a range-dependent correction
factor that accounts for the fact that the critical amplitude for small~$R$
deviates from the asymptotic range dependence. This only introduces a shift
along the vertical axis. The solid curve indicates the crossover function for
the block-shaped interaction profile and the dashed lines mark the mean-field
(``MF'') and the Ising asymptote. The numbers in the key refer to the values
for $R_1^2$ and~$R_2^2$ for each interaction profile.  For clarity not all
systems listed in Table~\protect\ref{tab:ranges} have been included in the
graph, but the omitted data points are fully compatible with those shown.}
\label{fig:chi}
\end{figure}

The quantity of central interest, however, is the magnetic
susceptibility~$\chi$ below~$T_{\rm c}$. In Fig.~\ref{fig:chi} the crossover
function for the block-shaped profile is shown as a reference curve. The
parameter along the horizontal axis is proportional to~$t/G$ ($G \propto
R^{-2d/(4-d)}$) and the susceptibility is divided by a factor~$R^2$ to obtain a
data collapse for different ranges.  One clearly observes how the solid curve
interpolates smoothly, but with a {\em non-monotonic\/} derivative, between the
Ising asymptote and the classical asymptote. Within the same figure, we have
also plotted the finite-size data for four different double-block profiles.
Several remarks are in order here.  First, all data have been divided by a
range-dependent factor describing the deviation of the connected susceptibility
from its asymptotic range dependence.  This is similar to the difference
between the dashed curve and the solid line in Fig.~\ref{fig:mabs}; as this
curve turns out to depend solely on the value of~$R$ and not on the {\em
shape\/} of the interaction profile, it is permissible to use the same
expression for all systems.  For large interaction ranges, the correction
factor approaches unity.  Secondly, at the right-hand side of the graph the
data points start to deviate from the reference curve.  This is caused by the
fact that, sufficiently close to~$T_{\rm c}$, the diverging correlation length
is truncated by the finite system size. For the systems with $R_1^2=4$
and~$R_1^2=93$ (both with $R_2^2=140$) this happens in the figure at different
temperatures, despite the fact that they have very similar values for the
effective range~$R$. The reason for this is that the data points pertain to
different system sizes, {\em viz.\/}\ $L=1000$ and~$L=300$, respectively. The
inset shows that for the {\em same\/} system size ($L=300$) the data points for
both systems virtually coincide, even in the finite-size regime!  Finally, the
left-most data points have been corrected for saturation effects, which are
fully described by mean-field theory, cf.\ Ref.~\cite{cross}.  This is merely
an optical issue: Also the saturated curves (which display a strong decrease of
the susceptibility) show no dependence on the shape of the interaction profile.
The primary message, however, of Fig.~\ref{fig:chi} is that for {\em all\/}
interaction profiles the data in the thermodynamic limit perfectly coincide
with the reference curve for the block-shaped potential.  We view this as a
strong indication that crossover functions possess a considerable degree of
universality and conclude that a second length scale ($R_1$) of the form
introduced in this work is insufficient to induce modifications of the
crossover functions, contrary to some expectations.

\section{Conclusions}
In summary, we have examined the critical properties of two-dimensional
Ising-like models with an extended interaction range, for several different
shapes of the interaction profile. In addition, we have calculated the
crossover function for the susceptibility in the low-temperature regime,
describing the crossover from classical to non-classical critical behaviour
upon approach of the critical point. Although recent work has suggested that
this function cannot be described in terms of a single parameter, namely the
reduced temperature divided by the Ginzburg number, we find that it is
independent of the precise shape of the interaction profile. Irrespective of
the presence of an additional length scale or a high asymmetry in the
interaction profile, all examined systems can be classified according to a
single additional parameter describing the effective interaction range. In
particular, the non-monotonic crossover of the effective susceptibility
exponent, as found in Ref.~\cite{chicross}, is not a peculiarity of the
block-shaped interaction profile, but can be observed for all systems studied
in the present work, provided that the effective interaction range is
sufficiently large.  Furthermore, a corollary of the results presented here is
that the coefficient~$u$ in the LGW Hamiltonian also appears to have only a
weak dependence on the {\em shape\/} of the interaction profile. Of course,
this still leaves the possibility that other parameters, that still need to be
identified, have a more pronounced influence on~$u$ and thus on the crossover
functions.

\stars 

E. Luijten gratefully acknowledges illuminating discussions with M. A.
Anisimov and J. V. Sengers. We thank the HLRZ J\"ulich for computing time on a
Cray-T3E.

\vskip -12pt

\end{document}

%% file: euromacr.tex
\def\And{{\rm and\ }}

\def\stars{\bigskip\centerline{***}\medskip}

\newif\ifboo \boofalse


%% file: dblock.bbl
\begin{thebibliography}{10}

\bibitem{nicoll81}
Nicoll J.~F. and Bhattacharjee J.~K., Phys. Rev. B, {\bf 23} (1981) 389.

\bibitem{bagnuls85}
Bagnuls C. and Bervillier C., Phys. Rev. B, {\bf 32} (1985) 7209.

\bibitem{bagnuls87}
Bagnuls C., Bervillier C., Meiron D.~I. and Nickel B.~G., Phys. Rev. B, {\bf
  35} (1987) 3585.

\bibitem{pelissetto98}
Pelissetto A., Rossi P. and Vicari E., Phys. Rev. E, {\bf 58} (1998) 7146.

\bibitem{belyakov92}
Belyakov M.~Y. and Kiselev S.~B., Physica A, {\bf 190} (1992) 75.

\bibitem{medran}
Luijten E., Bl{\"o}te H.~W.~J. and Binder K., Phys. Rev. E, {\bf 54} (1996)
  4626.

\bibitem{anisimov95}
Anisimov M.~A., Povodyrev A.~A., Kulikov V.~D. and Sengers J.~V., Phys. Rev.
  Lett., {\bf 75} (1995) 3146.

\bibitem{chi3d}
Luijten E. and Binder K., Phys. Rev. E, {\bf 58} (1998) R4060.

\bibitem{anisimov98}
Anisimov M.~A., Luijten E., Agayan V.~A., Sengers J.~V. and Binder K., preprint
  cond-mat/9810252.

\bibitem{anisimov92}
Anisimov M.~A., Kiselev S.~B., Sengers J.~V. and Tang S., Physica A, {\bf 188}
  (1992) 487.

\bibitem{chen90b}
Chen Z.~Y., Abbaci A., Tang S. and Sengers J.~V., Phys. Rev. A, {\bf 42} (1990)
  4470.

\bibitem{melnichenko97}
Melnichenko Y.~B., Anisimov M.~A., Povodyrev A.~A., Wignall G.~D., Sengers
  J.~V. and Van~Hook W.~A., Phys. Rev. Lett., {\bf 79} (1997) 5266.

\bibitem{kouvel64}
Kouvel J.~S. and Fisher M.~E., Phys. Rev., {\bf 136} (1964) A1626.

\bibitem{chicross}
Luijten E., Bl{\"o}te H.~W.~J. and Binder K., Phys. Rev. Lett., {\bf 79} (1997)
  561.

\bibitem{cross}
Luijten E., Bl{\"o}te H.~W.~J. and Binder K., Phys. Rev. E, {\bf 56} (1997)
  6540.

\bibitem{lr-alg}
Luijten E. and Bl{\"o}te H.~W.~J., Int. J. Mod. Phys. C, {\bf 6} (1995) 359.

\bibitem{q-2d}
Kamieniarz G. and Bl{\"o}te H.~W.~J., J. Phys. A, {\bf 26} (1993) 201.

\end{thebibliography}
